\documentstyle[preprint,floats,aps,pre]{revtex}
\begin{document}
\draft

\title{Asymmetric motion in a double-well under the action of
zero-mean Gaussian white noise and periodic forcing}

\author{Mangal C. Mahato and A.M. Jayannavar\\
Institute of Physics, Sachivalaya Marg,
Bhubaneswar-751005, India\\}


\maketitle

\begin{abstract}
Residence times of a particle in both the wells of a double-well
system, under the action of zero-mean Gaussian white noise and
zero-averaged but temporally asymmetric periodic forcings, are
recorded in a numerical simulation. The difference between the
relative mean residence times in the two wells shows monotonic
variation as a function of asymmetry in the periodic forcing and
for a given asymmetry the difference becomes largest at an
optimum value of the noise strength. Moreover, the passages from
one well to the other become less synchronous at small noise
strength as the asymmetry parameter (defined below) differs from
zero, but at relatively larger noise strengths the passages
become more synchronous with asymmetry in the field sweep. We
propose that asymmetric periodic forcing (with zero mean) could
provide a simple but sensible physical model for unidirectional
motion in a symmetric periodic system aided by a symmetric
Gaussian white noise. 
\end{abstract}


Several physical models have recently been
proposed\cite{Magn,Bart,Astu,Doer,Mill,Jaya,MCM,Pros} to
understand possible average asymmetric motion of a Brownian
particle in a periodic potential. Living systems are manifestly
nonequilibrium and quite understandably such an asymmetric
motion has been observed recently in biological
systems\cite{Svob}. Though the quest for extracting useful work
out of nonequilibrium systems is not new, the biological
experimental observation has given enough motivation recently to
renew effort in that direction. It has resulted in better
understanding of the problem and also it has helped in inventing
new devices for practical use\cite{Pros}. In the present work we
study a symmetric two-well system subjected to zero-mean Gaussian
white noise. We apply an external field that is periodic in
time. The external field is taken to be temporally asymmetric
but with mean force zero in a period. We find that eventhough
the mean deterministic force experienced by a particle due to
the external field is zero, the Gaussian white noise (centered
at zero) helps it to extract work while moving on a potential
surface. This is reflected in the asymmetric passages between
the two symmetrically connected wells modulated periodically by
the external field.  

We consider the symmetric two-well potential represented by
$U(m)=-\frac{a}{2}m^2+\frac{b}{4}m^4$ and consider the external
field $h(t)$ to be periodic in time and assume an asymmetric
saw-tooth form for it. The asymmetry in the saw-tooth form comes
because the positive and negative slopes are taken to have
different magnitude. The mean force in a period (because of the
external field) is assured to be zero by taking the maxima and
minima of the saw-tooth to have values $+h_0$ and $-h_0$,
respectively. A Brownian particle will thus experience a
combined (time dependent) potential
\begin{equation}
\Phi (m,t)=U(m)-mh(t).
\end{equation}
We consider the time evolution of the particle coordinate $m(t)$
to be governed by the overdamped Langevin equation
\begin{equation}
\dot m=-\frac{\partial \Phi}{\partial m}+\hat f(t),
\end{equation}
where $\hat f(t)$ is a randomly fluctuating force and is taken
to be Gaussian with statistics
\begin{equation}
<\hat f(t)>=0,
\end{equation}
and
\begin{equation}
<\hat f(t)\hat f(t')>=2D\delta (t-t').
\end{equation}
Here $<...>$ represents average over a large number of
realizations of the random forces.

Our calculation involves solving the Langevin equation
numerically and to monitor the time evolution of $m(t)$ for a
long time for given noise strength $D$. The calculation is done
for a fixed subcritical $h(t)$ with amplitude $h_0<h_c$, where
$h_c$ is the minimum value of $\vert h(t)\vert$ at which one of
the two wells of $\Phi (m)$ becomes unstable and disappears.
Since $h(t)<h_c$ always, the barrier between the two wells never
vanishes. Therefore to pass from one well to the other a
particle need necessarily have to surmount a nonzero potential
barrier and therefore has to be noise aided. Along the time axis
we record the events of passages between the two wells. We
consider passage to take place from a given well as and when the
trajectory $m(t)$, emerging from the given well, crosses the
inflexion point on the other side of the potential barrier
separating the two wells. From the markers recorded on the time
axis we obtain the distribution $\rho _1(\tau)$ of residence
times, $\tau$, in the well 1 and (similarly for the well 2). And
also the distribution $\rho _{12}(h)$ of field values $h(t)$ at
which passages take place from the well 1 to the well 2 (and
similarly from the well 2 to the well 1) are calculated from the
same recordings.  

The distributions $\rho _{12}(h)$ and $\rho _{21}(h)$ determine
the evolution of the fraction of population in a well as the
external field $h(t)$ varies. For example, the fraction $m_2(h)$
of the population in the well 2 evolves [from $(n-1)$th step
to $n$th step] as
\begin{equation}
m_2(n)=m_2(n-1)-m_2(n-1)\rho _{21}(h_{n-1})(h_n-h_{n-1})
+m_1(n-1)\rho _{12}(h_{n-1})(h_n-h_{n-1}),
\end{equation}
where $h_n$ is the field value at the $n$th subdivision point in
a cycle of $h$. The interval of uniform subdivisions
$(h_n-h_{n-1})$ are taken to be optimally small for better
accuracy. In our calculation we take $(h_n-h_{n-1})=\Delta
h=0.001h_c$. So the whole period is divided into
$N=\frac{2h_0}{\Delta h}$ equal segments. This evolution equation
together with the periodicity condition, for instance,
$m_2(n=0)=m_2(n=\frac {2h_0}{\Delta h})$ gives the closed
hysteresis loop $\bar m(h)=m_2(h)-m_1(h)$. Also, throughout our
calculation, we take $a=2.0$, and $b=1.0$. The hysteresis loop
area is a good measure of degree of synchronization of passages
between the two wells. For example, if the passages take place
only when the potential barrier for passage is the least, {\it i.e.,}
at $h=\pm h_0$, the distributions $\rho _{12}(h)$ and $\rho _{21}(h)$
will be sharply peaked at $h=h_0$ and $h=-h_0$, respectively. In
this case the hysteresis loop will be nearly rectangular and
therefore will have the largest area. On the other hand, if the 
passages take place all over and randomly (corresponding to the
case of least synchronization) so that $\rho _{12}(h)$
and $\rho _{21}(h)$ are uniform the loop area becomes the least.
We explore how the hysteresis loop area $A$ changes as a function
of the asymmetry of the field $h(t)$ and also as a function of
the noise strength $D$. The asymmetry of field sweep, $\Delta$,
is defined as $\Delta=\frac{T_1-\frac{T_0}{2}}{\frac{T_0}{2}}$,
where $T_1$ is the time for the field to change from $h_0$ to
$-h_0$ and $T_0$ is the period of oscillation of $h(t)$. Fig. 1
shows a typical hysteresis loop with $\Delta\ne 0$ as compared
to one with $\Delta=0$. Notice that the hysteresis loops do not
saturate for the field amplitude $h_0=.7h_c$ considered here.

Figure 2 shows the variation of hysteresis loop area $A$ as a
function of the asymmetry parameter $\Delta$ for fixed values of
$h_0$, $T_0$, and various values of $D$. The area is the largest
at $\Delta =0$ for small $D$ but is the lowest for relatively
larger $D$. In both the cases, however, $A(-\Delta)=A(\Delta)$.
This result is to be compared with our earlier work on two-well
systems\cite{MCM} where {\it first-passage times} instead of
residence times were calculated. In that work the upper half of
the hysteresis loop, corresponging to passages from well 2 to
well 1, was obtained from the first-passage-time distribution
$\rho (h)$ that spread between $h_0$ and $-h_0$ (and not over
the whole period) and the other half was obtained by symmetry.
Consequently the hysteresis loops, by construction, saturated to
$\bar m(h_0)=1.0h_c$ and $\bar m(-h_0)=-1.0h_c$, and were
symmetric for all $h_0$ including $h_0=.7h_c$. The variation of
hysteresis loop area, however, showed asymmetry and attained a
peak (usually not at $\Delta=0$) as a function of $\Delta$. In
Fig. 3, we show how, in the present work, area changes as a
function of $D$ for fixed $\Delta$, $T_0$, and $h_0$. However,
before discussing the physical significance of these results we
consider the mean residence times $\bar\tau _1$ and $\bar\tau
_2$ in the two wells as a function of $\Delta$.  

From the distributions $\rho _1(\tau)$ and $\rho _2(\tau)$ of
the residence times we calculate the mean residence times in
each of the two wells. We, then, calculate the fraction of
times, $f_1$ and $f_2$, the particle spends, on the average, in
the two wells. The difference, $M=f_2-f_1$, gives a quantity
analogous to magnetization (normalized) in magnetic systems. In
Fig. 4 we plot $M$ as a function of $\Delta$. From the figure it
is clear that $M(\Delta)=-M(-\Delta)$ (upto the order of
accuracy of our numerical calculation) and vary roughly
monotonically. This indicates that the particles will tend to
accumulate in the well 2 if $\Delta>0$ and in the well 1 if
$\Delta<0$. This conclusion is plausible because in the
situation under consideration the mean residence times
$\bar\tau_1$ and $\bar\tau_2$ add up to
$\bar\tau=\bar\tau_1+\bar\tau_2$ which is larger than $T_0$, the
period of oscilation of $h(t)$. This simply indicates that
passages do not take place in every cycle of $h(t)$. For
$\Delta>0$, for instance, there will be larger number of
passages from well 1 to 2 than from 2 to 1 per cycle of $h(t)$,
in an ensemble. This results in a net accumulation of particles
in the well 2, asymptotically. This asymmetry in passages is
also reflected indirectly in the hysteretic property of the
system. For $\Delta\ne 0$ the hysteresis loops are asymmetric.
Also, as $\Delta$ deviates from zero the hysteresis loop area
decreases for small $D$ (but increases for relatively larger
$D$) [Fig. 2], which indicates that now the passages are
less(more) synchronous with the input signal $h(t)$. This effect
indicates as if a net average constant field is applied in a
direction determined by $\Delta$. 

The net accumulation $M$ in a two-well system for given $\Delta$
changes with the noise strength $D$. Fig. 5 shows that $M$
increases initially, reaches a maximum, and then decreases
gradually as $D$ is increased. We, thus, have an optimum value
of $D$ at which the accumulation in the well 2 (when $\Delta>0$)
is the largest after a large number of cycles of $h(t)$. All the
results obtained in the present work are susceptible to
experimental verification by the recently developed optical
interferometric techniques\cite{Simo}.

All the calculated results discussed so far are valid for a
double-well potential. However, it is not difficult to envisage
the situation in case of a periodic potential. In a two-well
potential, for $\Delta>0$, as mentioned earlier, the number of
passages taking place from well 1 to well 2 per cycle of field
sweep is larger than the number of passages from well 2 to well
1. One may justifiably extrapolate this result to state that in
a periodic potential (that may even be symmetric), in a given
number of cycles of $h(t)$ there will be more $1\rightarrow 2$
passages than $2\rightarrow 1$ passages, and hence there will be
a net current of particles in the right direction $(1\rightarrow
2)$ when $\Delta>0$ and in the reverse direction when
$\Delta<0$. This current will increase with the magnitude of
$\Delta$. However, for a given $\Delta$ we can find an optimum
value of the Gaussian white noise strength $D$ at which the
current will be maximum. This is an important observation
because here we have a physical model for unidirectional motion
of a particle in a nonratchetlike symmetric periodic potential
aided by symmetric Gaussian white noise (fluctuating forces).

\vfil
\newpage
\begin{figure}
\caption{Hysteresis loops $\bar m(h)$, for $h_0=0.7h_c$,
$D=0.15$, and $T_0=28.0$, are plotted for (a) $\Delta=0$ (solid
line) and (b) $\Delta=0.5$ (dotted line).} 
~~~~~

\caption{Plots of hysteresis loop area $A$ versus
$\Delta$ for (a) $D=0.1 (\circ)$, (b) $D=0.15 (\Box)$, and (c)
$D=0.2 (\diamond)$, (d) $D=0.5 (\bigtriangleup)$, and (e) $D=0.7
(\bigtriangledown)$.} 
~~~~~

\caption{Plot of hysteresis area as a function of $D$
for (a) $\Delta=0.5 (\circ)$, and (b) $\Delta=\frac{13}{14}
(\Box)$.} 
~~~~~

\caption{Shows the variation of accumulation $M$ in a
well as a function of $\Delta$ for (a) $D=0.15 (\Box)$, (b)
$D=0.2 (\diamond)$, (c) $D=0.5 (\bigtriangleup)$, and (d) $D=0.7
(\bigtriangledown)$.} 
~~~~~

\caption{Plot of $M$ versus $D$ for (a) $\Delta=0.5
(\circ)$, and (b) $\Delta=\frac{13}{14} (\Box)$.}
\end{figure}


\begin{thebibliography}{999}
%
\bibitem{Magn}
M. Magnasco, {\it Phys. Rev. Lett.} {\bf 71}, 1477(1993).
%
\bibitem{Bart}
R. Bartussek, P. H\"anggi, and J.G. Kissner, {\it Europhys.
Lett.} {\bf 28}, 459(1994); R. Bartussek, P. Reimann, and P.
H\"anggi, {\it Phys. Rev. Lett.} {\bf 76}, 1166(1996); P.
H\"anggi, R. Bartussek, P. Talkner, and J. Luczka, {\it
Europhys. Lett.} {\bf 35}, 315(1996); P. Jung, J.K. Kissner, and
P. H\"anggi, {\it Phys. Rev. Lett.} {\bf 76}, 3436(1996).
%
\bibitem{Astu}
R.D. Astumian and M. Bier, {\it Phys. Rev. Lett.} {\bf 72},
1776(1994). 
%
\bibitem{Doer}
C.R. Doering, N. Horsethemke, and J. Riordan, {\it Phys. Rev.
Lett.} {\bf 72}, 2984(1994).
%
\bibitem{Mill}
M.M. Millonas, and M.I. Dykman, {\it Phys. Lett.} A{\bf 185},
65(1994); M.M. Millonas, {\it Phys. Rev. Lett.} {\bf 74},
10(1995). 
%
\bibitem{Jaya}
A.M. Jayannavar, {\it Phys. Rev.} E{\bf 53}, 2957(1996); M.C.
Mahato, T.P. Pareek, and A.M. Jayannavar {\it Int. J. Mod.
Phys.} {\bf B}, (in Press). 
%
\bibitem{MCM}
M.C. Mahato, and A.M. Jayannavar, {\it Phys. Lett.} A{\bf 209},
21(1995); D.R. Chialvo, and M.M. Millonas, {\it ibid} A{\bf
209}, 26(1995); For motion induced by high amplitude ($h_0>h_c$)
deterministic {\it ac} forces see A. Ajdari, D. Mukamel, L.
Peliti, and J. Prost, {\it J. Phys. I France} {\bf 4},
1551(1994). 
%
\bibitem{Svob}
K. Svoboda, C.F. Schmitt, B.J. Schnapp, and S.M. Block, {\it
Nature} {\bf 365}, 721(1993).
%
\bibitem{Pros}
J. Prost, J.F. Chauwin, L. Peliti, and A. Ajdari, {\it Phys.
Rev. Lett.} {\bf 72}, 2652(1994); J. Rousselet, L. Salome, A.
Ajdari, and J. Prost, {\it Nature} {\bf 370}, 446(1994); J.F.
Chauwin, A. Ajdari, and J. Prost, {\it Europhys. Lett.} {\bf
32}, 373(1995).
%
\bibitem{Simo}
A. Simon, and A. Libchaber, {\it Phys. Rev. Lett.} {\bf 68},
3375(1992). 
\end{thebibliography}
\end{document}